\documentclass[aps,prl,reprint,groupedaddress]{revtex4-1}
\usepackage{amsmath}
\usepackage{array}
\usepackage{multirow,bigdelim}
\usepackage{slashbox}

\begin{document}

\title{Method for classifying multi-qubit states via the rank of coefficient matrix and its application to four-qubit states
}
\author{Xiangrong Li$^1$, Dafa Li$^{2,3}$}
\address{$^1$ Department of Mathematics, University of
California, Irvine, CA 92697-3875, USA \\
$^2$ Department of mathematical sciences, Tsinghua University,
Beijing 100084 CHINA\\
$^3$ Center for Quantum Information Science and Technology,
Tsinghua National Laboratory for information science and technology
(TNList), Beijing 100084, CHINA}

\begin{abstract}
We construct
coefficient matrices of size $2^{\ell}$ by $2^{n-\ell}$ associated with pure 
$n$-qubit states and prove the invariance of the ranks of the 
coefficient matrices under stochastic local operations and classical
communication (SLOCC). 
The ranks give rise to a simple way of partitioning pure $n$-qubit states into
inequivalent families and distinguishing degenerate families from one
another under SLOCC.
Moreover, the classification scheme via the ranks of coefficient matrices can 
be combined with other schemes to build a more refined classification scheme. 
To exemplify we classify the nine families of four qubits
introduced by Verstraete \emph{et al.} [Phys. Rev. A 65, 052112 (2002)] further 
into inequivalent subfamilies via the ranks of coefficient matrices, 
and as a result, we find 28 genuinely entangled 
families and all the degenerate classes can be distinguished up to 
permutations of the four qubits. We also discuss the completeness of 
the classification of four qubits into nine families.
\end{abstract}

\maketitle

\section{I. Introduction}

Quantum entanglement plays a crucial role in quantum information theory,
with applications to quantum teleportation, quantum cryptography, and quantum
computation \cite{Nielsen}. The equivalence under stochastic local
operations and classical communication (SLOCC) induces a natural partition
of quantum states. The central task of SLOCC classification is to classify
quantum states according to a criterion that is invariant under SLOCC.

SLOCC entanglement classification has been the subject of intensive study
during the last decade \cite%
{Dur,Verstraete,Miyake,Piza,Chterental,Cao,LDF07b,Lamata,LDFEPL,LDFQIC09,Bastin,Borsten,Ribeiro,LDFQIC11,LDFJPA,Viehmann,LDF12b,Sharma,Buniy}%
. For three qubits, there are six SLOCC equivalence classes of which two are
genuinely entanglement classes: GHZ and $W$ \cite{Dur} and four degenerate
classes can be distinguished by the local ranks (i.e., ranks of
single-qubit reduced density matrices obtained by tracing out all but one
qubit \cite{Dur}).
For four or more qubits, there are infinite SLOCC classes and it is highly
desirable to partition the infinite classes into a finite number of
families. The key lies in finding criteria to determine which family an
arbitrary quantum state belongs to. In a pioneering work, Verstraete \emph{et al.}
\cite{Verstraete} obtained nine SLOCC inequivalent families of four qubits using
Lie group theory: $G_{abcd}$, $L_{abc_{2}}$, $L_{a_{2}b_{2}}$, $L_{ab_{3}}$,
$L_{a_{4}}$, $L_{a_{2}0_{3\oplus {\bar{1}}}}$, $L_{0_{5\oplus {\bar{3}}}}$, $%
L_{0_{7\oplus {\bar{1}}}}$, and $L_{0_{3\oplus {\bar{1}}}0_{3\oplus {\bar{1}}%
}}$. 
It is clear that, some families obtained by Verstraete \emph{et al.} \cite{Verstraete}
contain an infinite number of SLOCC classes and some contain both degenerate
classes and genuinely entangled classes. It is of great importance to find a
more refined partition of four-qubit states such that the degenerate classes
are distinguished from the genuinely entangled families.
Many other efforts have been devoted to the SLOCC
entanglement classification of four qubits \cite%
{Miyake,Chterental,Cao,LDF07b,Lamata,LDFQIC09,Borsten,Viehmann,Buniy}.
More recently, a few attempts have been made toward the generalization to
higher number of qubits, including odd $n$ qubits \cite{LDFQIC11},
even $n$ qubits \cite{LDFJPA}, symmetric $n$ qubits \cite%
{LDFEPL,Bastin,Ribeiro}, and general $n$ qubits \cite{LDF12b,Sharma}.

This paper is organized as follows. We first construct coefficient matrices
of size $2^{\ell}$ by $2^{n-\ell}$ associated to pure 
$n$-qubit states and prove the invariance of 
the ranks of coefficient matrices under SLOCC in Section II.
In Section III, we present a recursive formula which allows us to easily 
calculate the ranks of coefficient matrices of $n$-qubit biseparable states. 
We next show that the degenerate families of general $n$ qubits
are inequivalent to one another under SLOCC in Section IV.
Section V is devoted to the classification of four qubits via the ranks of 
coefficient matrices.
Section VI provides the discussion of the completeness of the nine families
obtained by Verstraete \emph{et al.} \cite{Verstraete}.
We finally conclude this paper in Section VII.

\section{II. The invariance of the ranks of coefficient matrices}


Let $|\psi \rangle_{1\cdots n}=\sum_{i=0}^{2^{n}-1}a_{i}|i\rangle $ be an $n$%
-qubit pure state. We associate with the state $|\psi
\rangle_{1\cdots n}$ a $2^{\ell}$ by the $2^{n-\ell}$ coefficient matrix $%
C_{1\cdots \ell,(\ell+1)\cdots n}(|\psi \rangle_{1\cdots n})$ whose entries
are the coefficients $a_{0},a_{1},\cdots ,a_{2^{n}-1}$ of the state $|\psi
\rangle_{1\cdots n}$ arranged in ascending lexicographical order. To
illustrate, we list $C_{1\cdots \ell,(\ell+1)\cdots n}(|\psi
\rangle_{1\cdots n})$ below as:
\begin{equation}
\left(
\begin{array}{cccc}
a_{\underbrace{0\cdots 0}_{\ell}\underbrace{0\cdots 0}_{n-\ell}} & \cdots &
a_{\underbrace{0\cdots 0}_{\ell}\underbrace{1\cdots 1}_{n-\ell}} &  \\
a_{\underbrace{0\cdots 1}_{\ell}\underbrace{0\cdots 0}_{n-\ell}} & \cdots &
a_{\underbrace{0\cdots 1}_{\ell}\underbrace{1\cdots 1}_{n-\ell}} &  \\
\vdots & \vdots & \vdots &  \\
a_{\underbrace{1\cdots 1}_{\ell}\underbrace{0\cdots 0}_{n-\ell}} & \cdots &
a_{\underbrace{1\cdots 1}_{\ell}\underbrace{1\cdots 1}_{n-\ell}} &
\end{array}%
\right) .  \label{ma-1}
\end{equation}

In the binary form of the coefficient matrix in Eq. (\ref{ma-1}), bits $1$
to ${\ell}$ and $\ell+1$ to $n$ are referred to as the row bits and column
bits, respectively. If $\ell=0$, $C_{\emptyset,1\cdots n}(|\psi
\rangle_{1\cdots n})$ reduces to the row vector $(a_0,\cdots, a_{2^n-1})$,
and if $\ell=n$, $C_{1\cdots n,\emptyset}(|\psi \rangle_{1\cdots n})$
reduces to the column vector $(a_0,\cdots, a_{2^n-1})^T$.

Let $\{q_{1},q_{2},\cdots ,q_{n}\}$ be a permutation of $\{1,2,\cdots ,n\}$.
Let $C_{q_{1}\cdots q_{\ell},q_{\ell+1}\cdots q_{n}}(|\psi \rangle_{1\cdots
n})$ be the $2^{\ell}\times 2^{n-\ell}$ coefficient matrix of the state $%
|\psi \rangle_{1\cdots n}$, which is constructed from the coefficient
matrix $C_{12\cdots \ell,\ell+1\cdots n}$ in Eq. (\ref{ma-1}) by taking the
corresponding permutation. Here $q_1,\cdots,q_\ell$ are the row bits and $%
q_{\ell+1},\cdots,q_n$ are the column bits. Indeed, we only need to specify
the row bits, as the column bits would simply be the rest of the bits. In
the sequel, we will omit the subscripts $q_{\ell+1},\cdots,q_n$ and simply
write $C_{q_{1}\cdots q_{\ell}}$, whenever the column bits are clear from
the context.

It is known that two $n$-qubit pure states $|\psi \rangle_{1\cdots n}$ and $%
|\psi^{\prime }\rangle_{1\cdots n}$ are equivalent to each other under SLOCC
if and only if there are local invertible operators $\mathcal{A}_{1}$, $%
\mathcal{A}_{2},\cdots $, and $\mathcal{A}_{n}$ such that \cite{Dur}
\begin{equation}
|\psi ^{\prime }\rangle_{1\cdots n}=\mathcal{A}_{1}\otimes \mathcal{A}%
_{2}\otimes \cdots \otimes \mathcal{A}_{n}|\psi \rangle_{1\cdots n}.
\label{slocc-eq}
\end{equation}

In terms of coefficient matrices, it can be verified that the following
result holds: For any two SLOCC equivalent $n$-qubit pure states $|\psi
\rangle_{1\cdots n}$ and $|\psi^{\prime }\rangle_{1\cdots n}$, their
coefficient matrices $C_{q_{1}\cdots q_{\ell}}$ satisfy the equation:
\begin{eqnarray}
&&C_{q_{1}\cdots q_{\ell}}(|\psi ^{\prime }\rangle_{1\cdots n})=\notag \\
&&(\mathcal{A}_{q_{1}}\otimes \cdots \otimes \mathcal{A}_{q_{\ell}}) 
C_{q_{1}\cdots q_{\ell}}(|\psi \rangle_{1\cdots n}) (\mathcal{A}%
_{q_{\ell+1}}\otimes \cdots \otimes \mathcal{A}_{q_{n}})^{T},\notag \\
\label{reduce-2}
\end{eqnarray}
where $\mathcal{A}_{1}, \mathcal{A}_{2}, \cdots$, and $\mathcal{A}_{n}$ are
the local operators in Eq. (\ref{slocc-eq}). Conversely, if there are local
invertible operators $\mathcal{A}_{1}, \mathcal{A}_{2}, \cdots$, and $%
\mathcal{A}_{n}$ such that Eq. (\ref{reduce-2}) holds true for some $%
C_{q_{1}\cdots q_{\ell}}$, then $|\psi \rangle_{1\cdots n}$ and $%
|\psi^{\prime }\rangle_{1\cdots n}$ are equivalent under SLOCC.

It immediately follows from Eq. (\ref{reduce-2}) that the rank of any
coefficient matrix of an $n$-qubit pure state is invariant under SLOCC. This
leads to the following theorem.

\textsl{Theorem 1.} If two $n$-qubit pure states are SLOCC equivalent then
their coefficient matrices $C_{q_{1}\cdots q_{\ell}}$ given above have the
same rank.

Restated in the contrapositive the theorem reads: If two coefficient
matrices $C_{q_{1}\cdots q_{\ell}}$ associated with two $n$-qubit pure
states differ in their ranks, then the two states belong necessarily to
different SLOCC classes.

Coefficient matrices constructed above turn out to be closely related to
reduced density matrices. We let $\rho _{12\cdots n}(|\psi \rangle_{1\cdots
n}) =|\psi \rangle _{1\cdots n}{}_{1\cdots n}\langle \psi |$ be the density
matrix of an $n$-qubit pure state $|\psi \rangle_{1\cdots n}$, and we let $%
\rho _{q_{1}\cdots q_{\ell}}$ be the $\ell$-qubit reduced density matrix
obtained from $\rho _{12\cdots n}$ by tracing out $n-\ell$ qubits. As has
been previously noted for bipartite systems of dimensions $d\times d$, a
reduced density matrix has a full rank factorization in terms of the
corresponding coefficient matrix and its conjugate transpose \cite{LDFCTP}.
This factorization also holds for $n$-qubit states \cite{Huang2}:
\begin{equation}
\rho _{q_{1}\cdots q_{\ell}}(|\psi \rangle_{1\cdots n}) =C_{q_{1}\cdots
q_{\ell}}(|\psi \rangle_{1\cdots n}) C_{q_{1}\cdots q_{\ell}}^{\dagger
}(|\psi \rangle_{1\cdots n}),  \label{full_rank}
\end{equation}%
where $C^{\dagger }$ is the conjugate transpose of $C$. An important
relationship between reduced density matrices and SLOCC polynomial
invariants can be obtained by taking the determinants of both sides of Eq. (%
\ref{full_rank}) for even $n$ and for $\ell=n/2$, yielding:
\begin{equation}  \label{eq_det}
\det \rho _{q_{1}\cdots q_{n/2}}(|\psi \rangle_{1\cdots n}) =\bigl|\det
C_{q_{1}\cdots q_{n/2}}(|\psi \rangle_{1\cdots n})\bigr|^2.
\end{equation}
Here $\det C_{q_{1}\cdots q_{n/2}}(|\psi \rangle_{1\cdots n})$ is a SLOCC
polynomial invariant of degree $2^{n/2}$ for even $n$ qubits and its
absolute value can be used as an entanglement measure \cite{LDFJPA12}. Thus
we have the following:

\textsl{Theorem 2.} For even $n$-qubit pure states, the determinants of $n/2$%
-qubit reduced density matrices are the squares of the SLOCC polynomial
invariants of degree $2^{n/2}$, with the absolute values of the latter
quantifying $n/2$-qubit entanglement of the even $n$-qubit states after
tracing out the other $n/2$ qubits.

As an example, when $n=4$ we have $\det \rho _{12}=|L|^{2}$, $\det \rho
_{13}=|M|^{2}$, and $\det \rho _{14}=|N|^{2}$, where $L$, $M$, and $N$ are
polynomial invariants of degree 4 \cite{Luque}. When $n=6$, there are 10
three-qubit reduced density matrices and 10 polynomial invariants of degree 8: $%
D_6^1,\cdots,D_6^{10}$ \cite{LDFJPA12}. For reduced density matrix $%
\rho_{123}$ and polynomial invariant $D_6^1$, we have $\det
\rho_{123}=|D_6^1|^2$. Similar equations hold for other reduced density
matrices and polynomial invariants with appropriate permutations of qubits.

\textsl{Remark 1.} (i). The determinants of reduced density matrices are
invariant under SLOCC. (ii). It is worth noting that Eq. (\ref{eq_det})
holds for bipartite systems of dimensions $d\times d$ as well \cite{LDFCTP}.

As a particular case of Eq. (\ref{full_rank}), when $q_i=i$ we have $\rho
_{1\cdots n}(|\psi \rangle_{1\cdots n}) =C_{1\cdots n}(|\psi
\rangle_{1\cdots n}) C_{1\cdots n}^{\dagger }(|\psi \rangle_{1\cdots n})$.
By virtue of Eq. (\ref{full_rank}), the rank of the $\ell$-qubit reduced density
matrix and the rank of the corresponding coefficient matrix are the same. In
light of Theorem 1, we have the following result.

\textsl{Corollary}. The ranks of $\ell$-qubit reduced density matrices
obtained by tracing out $n-\ell$ qubits are invariant under SLOCC.

This is particularly true for the local ranks \cite{Dur}. Note also that any
complex matrix has a singular value decomposition, with the number of
nonzero singular values equal to the rank of the matrix. This means that the
number of nonzero singular values of any coefficient matrix of an $n$-qubit
pure state is invariant under SLOCC.

\section{III. A recursive formula for the ranks of $N$-qubit biseparable states}

In principle, we can calculate the ranks of coefficient matrices for $n$%
-qubit biseparable pure states by direct calculations. However, in practice,
this is rather cumbersome from the computational point of view, and as $n$
becomes large, this might pose a serious problem. In order to avoid this
difficulty, we propose a simple recursive formula for the ranks of $n$-qubit
biseparable states.

Suppose that a biseparable $n$-qubit pure state $|\psi \rangle _{1\cdots n}$
is of the form $|\psi \rangle _{1\cdots n}=|\phi \rangle _{j_{1}\cdots
j_{k}}\otimes |\varphi \rangle _{j_{k+1}\cdots j_{n}}$ with $|\phi \rangle
_{j_{1}\cdots j_{k}}$ being a $k$-qubit state and $|\varphi \rangle
_{j_{k+1}\cdots j_{n}}$ being an $(n-k)$-qubit state. We let $C_{q_{1}\cdots
q_{\ell }}(|\psi \rangle _{1\cdots n})$ be the coefficient matrix associated
with the state $|\psi \rangle _{1\cdots n}$. We let $C_{q_{1}^{\ast }\cdots
q_{s}^{\ast }}(|\phi \rangle _{j_{1}\cdots j_{k}})$ be the $2^{s}$ by $%
2^{k-s}$ coefficient matrix associated with the $k$-qubit state $|\phi
\rangle _{j_{1}\cdots j_{k}}$. Here $\{q_{1}^{\ast },\cdots ,q_{s}^{\ast
}\}=\{q_{1},\cdots ,q_{\ell }\}\cap \{j_{1},\cdots ,j_{k}\}$ are the row
bits, and by convention, the rest $k-s$ bits are the column bits. Moreover,
we let $C_{q_{1}^{\prime }\cdots q_{t}^{\prime }}(|\varphi \rangle
_{j_{k+1}\cdots j_{n}})$ be the $2^{t}$ by $2^{n-k-t}$ coefficient matrix
associated with the $(n-k)$-qubit state $|\varphi \rangle _{j_{k+1}\cdots
j_{n}}$. Here $\{q_{1}^{\prime },\cdots ,q_{t}^{\prime }\}=\{q_{1},\cdots
,q_{\ell }\}\cap \{j_{k+1},\cdots ,j_{n}\}$ are the row bits, and by
convention, the rest $n-k-t$ bits are the column bits. It can be verified
that
\begin{align}
& C_{q_{1}\cdots q_{\ell }}(|\phi \rangle _{j_{1}\cdots j_{k}}\otimes
|\varphi \rangle _{j_{k+1}\cdots j_{n}})  \notag \\
& =C_{q_{1}^{\ast }\cdots q_{s}^{\ast }}(|\phi \rangle _{j_{1}\cdots
j_{k}})\otimes C_{q_{1}^{\prime }\cdots q_{t}^{\prime }}(|\varphi \rangle
_{j_{k+1}\cdots j_{n}}).
\end{align}%
In view of the fact that the rank of the Kronecker product of two matrices
is the product of their ranks, we arrive at the following recursive
formula for the ranks of coefficient matrices of an $n$-qubit biseparable
state:
\begin{align}
& \mbox{rank}(C_{q_{1}\cdots q_{\ell }}(|\phi \rangle _{j_{1}\cdots j_{k}}\otimes
|\varphi \rangle _{j_{k+1}\cdots j_{n}}))  \notag  \label{recursion} \\
& =\mbox{rank}(C_{q_{1}^{\ast }\cdots q_{s}^{\ast }}(|\phi \rangle _{j_{1}\cdots
j_{k}}))  
\mbox{rank}(C_{q_{1}^{\prime }\cdots q_{t}^{\prime }}(|\varphi \rangle
_{j_{k+1}\cdots j_{n}})).
\end{align}

The formula above allows us to calculate recursively the ranks of
coefficient matrices of $n$-qubit biseparable states in terms of the ranks
of coefficient matrices of $k$-qubit states and $(n-k)$-qubit states. To
illustrate the use of the recursive formula, we start with the initial
values $\mbox{rank}(C_{A}(|\phi \rangle _{A}))=1$ and $\mbox{rank}(C_{\emptyset}(|\phi
\rangle _{A}))=1$. It is known that a two-qubit pure state can be either of
the form $A$--$B$ (separable) or the form $AB$ (EPR). Using the recursive
formula, we find $\mbox{rank}(C_{A}(|\phi \rangle _{A}|\varphi \rangle
_{B}))=\mbox{rank}(C_{A}(|\phi \rangle _{A}))\times \mbox{rank}(C_{\emptyset }(|\varphi
\rangle _{B}))=1$. On the other hand, a direct calculation shows that $%
\mbox{rank}(C_{A}(|\varphi \rangle _{AB}))=2$. Using the results obtained above, we
can find the ranks of coefficient matrices of three-qubit pure states.
Consider, for example, $\mbox{rank}(C_{C}(|\phi \rangle _{B}|\varphi \rangle
_{AC})) $ for biseparable states being of the form $B$--$AC$. Using the
recursive formula, we have $\mbox{rank}(C_{C}(|\phi \rangle _{B}|\varphi \rangle
_{AC}))= \mbox{rank}(C_{\emptyset }(|\phi \rangle _{B})) \times \mbox{rank}(C_{C}(\varphi
\rangle _{AC}))=2$. In a similar fashion, we can fill in the rest of the
entries in Table \ref{table1}, except those in the last row which can be
obtained by direct calculations. Proceeding in this way, we can construct
Tables \ref{table2} and \ref{table3} for the ranks of coefficient matrices
for four and five qubits.

Note that in Tables \ref{table1} and \ref{table2} the ranks of only $%
2^{n-1}-1$ coefficient matrices are shown. This is due to the fact that
interchanging two row (resp. column) bits or exchanging the row and column
bits of a coefficient matrix does not alter the rank of the matrix, since
the former is equivalent to interchanging two rows (resp. columns) of the
matrix and the latter is equivalent to transposing the matrix. Ignoring $%
C_{\emptyset}$ and $C_{1\cdots n}$ which always have rank 1, this amounts to
totally $2^{n-1}-1$ potentially different coefficient matrices. For example,
the ranks of $C_{BA}$ and $C_{BC}$ are not shown in Table \ref{table2},
since $C_{AB}$ and $C_{BA}$ differ by the interchange of two rows, and $%
C_{BC}$ is the transpose of $C_{AD}$. As illustrated in Tables \ref{table1},
\ref{table2}, and \ref{table3}, the ranks of coefficient
matrices permit the partitioning of the space of the pure states into
inequivalent families under SLOCC (i.e., two states belong to the same
family if and only if the ranks of coefficient matrices are all equal). In
particular, degenerate families of three, four, and five qubits are
inequivalent from one another under SLOCC.

\begin{table}[tbph]
\caption{Ranks of coefficient matrices of three-qubit pure states.}
\label{table1}\renewcommand\arraystretch{1.5} \center
\begin{ruledtabular}
\begin{tabular}{llll}
\backslashbox[22mm]{{\scriptsize Families}}{{\scriptsize Ranks of}} & $C_{A}$ & $C_{B}$ & $C_{C}$
\\ \hline
$A$--$B$--$C$ & $1$ & $1$ & $1$ \\
$A$--$BC$ & $1$ & $2$ & $2$ \\
$B$--$AC$ & $2$ & $1$ & $2$ \\
$C$--$AB$ & $2$ & $2$ & $1$ \\
$ABC$ & $2$ & $2$ & $2$ \\
\end{tabular}
\end{ruledtabular}
\end{table}

\begin{table}[tbph]
\caption{Ranks of coefficient matrices of four-qubit pure states.}
\label{table2}\renewcommand\arraystretch{1.5} \center
\begin{ruledtabular}
\begin{tabular}{llllllll}
\backslashbox[12mm]{{\scriptsize Families}}{{\scriptsize Ranks of}} & $C_{A}$ & $C_{B}$ & $C_{C}$ &
$C_{D}$ & $%
C_{AB}$ & $C_{AC}$ & $C_{AD}$ \\ \hline
$A$--$B$--$C$--$D$ & $1$ & $1$ & $1$ & $1$ & $1$ & $1$ & $1$ \\
$A$--$B$--$CD$ & $1$ & $1$ & $2$ & $2$ & $1$ & $2$ & $2$ \\
$A$--$C$--$BD$ & $1$ & $2$ & $1$ & $2$ & $2$ & $1$ & $2$ \\
$A$--$D$--$BC$ & $1$ & $2$ & $2$ & $1$ & $2$ & $2$ & $1$ \\
$B$--$C$--$AD$ & $2$ & $1$ & $1$ & $2$ & $2$ & $2$ & $1$ \\
$B$--$D$--$AC$ & $2$ & $1$ & $2$ & $1$ & $2$ & $1$ & $2$ \\
$C$--$D$--$AB$ & $2$ & $2$ & $1$ & $1$ & $1$ & $2$ & $2$ \\
$A$--$BCD$ & $1$ & $2$ & $2$ & $2$ & $2$ & $2$ & $2$ \\
$B$--$ACD$ & $2$ & $1$ & $2$ & $2$ & $2$ & $2$ & $2$ \\
$C$--$ABD$ & $2$ & $2$ & $1$ & $2$ & $2$ & $2$ & $2$ \\
$D$--$ABC$ & $2$ & $2$ & $2$ & $1$ & $2$ & $2$ & $2$ \\
$AB$--$CD$ & $2$ & $2$ & $2$ & $2$ & $1$ & $4$ & $4$ \\
$AC$--$BD$ & $2$ & $2$ & $2$ & $2$ & $4$ & $1$ & $4$ \\
$AD$--$BC$ & $2$ & $2$ & $2$ & $2$ & $4$ & $4$ & $1$ \\
$ABCD^{\mbox{a}}$\footnotemark[0] & $2$ & $2$ & $2$ & $2$ & $\geq 2$ &
$\geq 2$ & $\geq 2$
\end{tabular}
\end{ruledtabular}
\footnotetext[0]{
$^{\mbox{a}}$ $ABCD$ can be further partitioned under SLOCC in terms of the
ranks of $C_{AB} $, $C_{AC}$ and $C_{AD}$.}
\end{table}

\begin{table}[tbph]
\caption{Ranks of coefficient matrices of five-qubit pure states.}
\label{table3}\renewcommand\arraystretch{1.5} \center
\begin{ruledtabular}
\begin{tabular}{lll}
\backslashbox{{\scriptsize Families}}{{\scriptsize Ranks of}} &
$C_{\alpha}$ & $C_{\beta \gamma (\beta\neq \gamma)}$ \\
\hline $i$--$j$--$k$--$\ell$--$m$ & 1$^{\mbox{b}}$ & 1$^{\mbox{c}}$
\\
$i$--$j$--$k$--$\ell m$ &
1, if $\alpha= i,j,k$ & 1, if $\beta, \gamma=i,j,k$ \\
& 2, otherwise & \quad or $\beta, \gamma=\ell,m$ \\
& & 2, otherwise \\ 
$i$--$jk$--$\ell m$ & 1, if $\alpha=i$ & 1, if $\beta, \gamma=j,k$ \\
& 2, otherwise & \quad or $\beta, \gamma=\ell,m$ \\
& & 2, if $\beta =i$ or $\gamma=i$ \\
& & 4, otherwise \\  
$i$--$j$--$k\ell m$ & 1, if
$\alpha=i$ or $j$ & 1, if
$\beta, \gamma=i,j$ \\
& 2, otherwise & 2, otherwise \\ 
$i$--$jk\ell m$ & 1, if $\alpha=i$ & 2, if $\beta=i$ or $\gamma=i$ \\
& 2, otherwise & 2, 3, or 4, otherwise \\ 
$ij$--$k\ell m$ & 2$^{\mbox{b}}$ & 1, if
$\beta, \gamma=i,j$ \\
& & 2, if $\beta, \gamma=k,\ell,m$ \\
& & 4, otherwise \\ 
$ijk\ell m$ & 2$^{\mbox{b}}$ & 2, 3, or 4$^{\mbox{c}}$
\end{tabular}
\end{ruledtabular}
\footnotetext[0]{$^{\mbox{a}}$
$\{i,j,k,\ell,m\}$ is any permutation of $\{A,B,C,D,E\}$. \newline
$^{\mbox{b}} $ $\alpha=i,j,k,\ell,m$. \newline
$^{\mbox{c}}$ $\beta, \gamma =i,j,k,\ell,m$.}%
\end{table}

\section{IV. Degenerate families of general $N$ qubits are SLOCC inequivalent to
one another}

The recursive formula above further gives rise to a criterion for
biseparability of an $n$-qubit pure state. Indeed, we note that Eq. (\ref%
{recursion}) holds particularly true for $\{q_{1},\cdots ,q_{\ell
}\}=\{j_{1},\cdots ,j_{k}\}$. In this case, the coefficient matrices $%
C_{q_{1}^{\ast }\cdots q_{s}^{\ast }}$ and $C_{q_{1}^{\prime }\cdots
q_{t}^{\prime }}$ reduce to a column vector and a row vector respectively,
and therefore both of them have rank 1. It follows that $\mbox{rank}(C_{q_{1}\cdots
q_{\ell }}(|\phi \rangle _{q_{1}\cdots q_{\ell }}\otimes |\varphi \rangle
_{q_{\ell +1}\cdots q_{n}}))=1$. Conversely, if $\mbox{rank}(C_{q_{1}\cdots q_{\ell
}}(|\psi \rangle _{1\cdots n}))=1$ for an $n$-qubit pure state $|\psi
\rangle _{1\cdots n}$, then $|\psi \rangle _{1\cdots n}$ is biseparable,
being of the form $|\psi \rangle _{1\cdots n}=|\phi \rangle _{q_{1}\cdots
q_{\ell }}\otimes |\varphi \rangle _{q_{\ell +1}\cdots q_{n}}$. This can be
seen as follows. For simplicity, we assume $q_{i}=i$ with $i=1,\cdots ,n$.
If $\mbox{rank}(C_{12\cdots \ell }(|\psi \rangle _{1\cdots n}))=1$, then all
columns of $C_{12\cdots \ell }$ are proportional to each other and each
column can be written into the form $(a_{0}b_{j},a_{1}b_{j},\cdots
,a_{2^{\ell }-1}b_{j})^{T}$. Hence, $|\psi \rangle _{1\cdots n}$ can be
written as $|\psi \rangle _{1\cdots n}=|\phi \rangle _{1\cdots \ell }\otimes
|\varphi \rangle _{(\ell +1)\cdots n}$ with $|\phi \rangle _{1\cdots \ell
}=\sum_{i=0}^{2^{\ell }-1}a_{i}|i\rangle _{1\cdots \ell }$ and $|\varphi
\rangle _{(\ell +1)\cdots n}=\sum_{j=0}^{2^{n-\ell }-1}b_{j}|j\rangle
_{(\ell +1)\cdots n}$. This leads to the following biseparability criterion
for $n$-qubit pure states.

\textsl{Biseparability criterion for $n$-qubit pure states}. For any
coefficient matrix $C_{q_{1}\cdots q_{\ell}}$ associated with an $n$-qubit
pure state $|\psi\rangle_{1\cdots n}$, $\mbox{rank}(C_{q_{1}\cdots q_{\ell}}(|\psi
\rangle_{1\cdots n}))=1$ if and only if $|\psi \rangle $ is biseparable,
being of the form $|\psi \rangle_{1\cdots n}=|\phi \rangle _{q_{1}\cdots
q_{\ell}}\otimes |\varphi \rangle _{q_{\ell+1}\cdots q_{n}}$ (see also \cite%
{Iwai,Huang2}).

Invoking the fact that an $n$-qubit pure state is entangled if it is not
full separable, we have the following criterion to identify $n$%
-qubit entangled (respectively, genuinely entangled) pure states: An $n$-qubit pure
state is entangled (respectively, genuinely entangled) if and only if the rank of at
least one of its coefficient matrices is (respectively, the ranks of its all
coefficient matrices are) greater than 1.

Note that all the above criteria can be rephrased in terms of the ranks of $%
\ell$-qubit reduced density matrices obtained by tracing out $n-\ell$ qubits
\cite{Chong} or the number of nonzero singular values of coefficient
matrices.

Theorem 1 together with the biseparability criterion above yield the
following theorem.

\textsl{Theorem 3}. Degenerate families of general $n$ qubits are
inequivalent to one another under SLOCC and they can be distinguished in
terms of the ranks of coefficient matrices (or in terms of the ranks of $%
\ell $-qubit reduced density matrices obtained by tracing out $n-\ell$
qubits).

The validity of Theorem 3 can be seen as follows. Given an $n$-qubit pure
state, a partition $P$ of the $n$ particles is a collection of disjoint sets
in such a way that the particles within any one set are entangled and any
two particles from different sets are not entangled. Suppose $F_{1}$ and $%
F_{2}$ are two different degenerate families with partitions $P_{1}$ and $%
P_{2}$ respectively. Without loss of generality, we assume that there exists
a set $S$ such that $S\in P_{1}$ and $S\not\in P_{2}$. Then the states in $%
F_{1}$ can be written in the biseparable form $|\phi \rangle _{S}|\varphi
\rangle _{\bar{S}}$, where ${\bar{S}}$ is the set of all particles except
those in $S$. According to the biseparability criterion above, $%
\mbox{rank}(C_{S})=1 $ for states in $F_{1}$. Since the states in $F_{2}$ cannot be
written in the above biseparable form, $\mbox{rank}(C_{S})>1$ for states in $F_{2}$%
. In light of Theorem 1, the two degenerate families are inequivalent to
each other under SLOCC.

In addition, we remark that degenerate families of general $n$ qubits can also
be distinguished from one another under SLOCC in terms of the ranks of $\ell
$-qubit reduced density matrices obtained by tracing out $n-\ell $ qubits or
the number of nonzero singular values of coefficient matrices.

\section{V. SLOCC classification of four qubits via the ranks of coefficient
matrices}

Suppose that the states $|\psi \rangle $ and $|\psi ^{\prime }\rangle $ of
four qubits are SLOCC equivalent to each other, then there are local
invertible operators $\mathcal{A}_{1}$, $\mathcal{A}_{2}$, $\mathcal{A}_{3}$%
, and $\mathcal{A}_{4}$ such that \cite{Dur}
\begin{equation}
|\psi ^{\prime }\rangle =\mathcal{A}_{1}\otimes \mathcal{A}_{2}\otimes
\mathcal{A}_{3}\otimes \mathcal{A}_{4}|\psi \rangle .  \label{slocc-eq-}
\end{equation}%
For a four-qubit state $|\psi \rangle =\sum_{i=0}^{15}a_{i}|i\rangle $, we
consider three coefficient matrices $C_{AB}$, $C_{AC}$, and $C_{AD}$ as
follows:
\begin{eqnarray}
C_{AB} &=&\left(
\begin{array}{cccc}
a_{0} & a_{1} & a_{2} & a_{3} \\
a_{4} & a_{5} & a_{6} & a_{7} \\
a_{8} & a_{9} & a_{10} & a_{11} \\
a_{12} & a_{13} & a_{14} & a_{15}%
\end{array}%
\right) , \\
C_{AC} &=&\left(
\begin{array}{cccc}
a_{0} & a_{1} & a_{4} & a_{5} \\
a_{2} & a_{3} & a_{6} & a_{7} \\
a_{8} & a_{9} & a_{12} & a_{13} \\
a_{10} & a_{11} & a_{14} & a_{15}%
\end{array}%
\right) , \\
C_{AD} &=&\left(
\begin{array}{cccc}
a_{0} & a_{4} & a_{2} & a_{6} \\
a_{1} & a_{5} & a_{3} & a_{7} \\
a_{8} & a_{12} & a_{10} & a_{14} \\
a_{9} & a_{13} & a_{11} & a_{15}%
\end{array}%
\right) .
\end{eqnarray}%
The coefficient matrices above satisfy the following equations:
\begin{eqnarray}
C_{AB}(|\psi ^{\prime }\rangle ) &=&\mathcal{A}_{1}\otimes \mathcal{A}%
_{2}C_{AB}(|\psi \rangle )(\mathcal{A}_{3}\otimes \mathcal{A}_{4})^{T},
\label{redu-1} \\
C_{AC}(|\psi ^{\prime }\rangle ) &=&\mathcal{A}_{1}\otimes \mathcal{A}%
_{3}C_{AC}(|\psi \rangle )(\mathcal{A}_{2}\otimes \mathcal{A}_{4})^{T},
\label{redu-2} \\
C_{AD}(|\psi ^{\prime }\rangle ) &=&\mathcal{A}_{1}\otimes \mathcal{A}%
_{4}C_{AD}(|\psi \rangle )(\mathcal{A}_{3}\otimes \mathcal{A}_{2})^{T}.
\label{redu-3}
\end{eqnarray}

It follows from Eqs. (\ref{redu-1})-(\ref{redu-3}) that
if two four-qubit states are SLOCC equivalent then their coefficient
matrices $C_{AB}$ (and also $C_{AC}$ and $C_{AD}$) have the same rank. Conversely,
if one of the coefficient matrices $C_{AB}$, $C_{AC}$, and $C_{AD}$ differ
in the ranks, then the two four-qubit states are SLOCC inequivalent. Let
family $F_{r_{AB}}^{C_{AB}}$ be the set of all four-qubit states with the
same rank $r_{AB}$ of the coefficient matrix $C_{AB}$. Here $r_{AB}$ ranges
over the values 1, 2, 3, and 4. Clearly, each one of the nine families
introduced by Verstraete \emph{et al.} \cite{Verstraete} can be further divided into
four SLOCC inequivalent subfamilies corresponding to the four possible
values of $r_{AB}$. In a similar manner, we can define the families $%
F_{r_{AC}}^{C_{AC}}$ and $F_{r_{AD}}^{C_{AD}}$. One can obtain a more
refined partition by further dividing the families $F_{r_{AB}}^{C_{AB}}$, $%
F_{r_{AC}}^{C_{AC}}$, and $F_{r_{AD}}^{C_{AD}}$ into subfamilies $%
F_{r_{AB}r_{AC}r_{AD}}^{C_{AB}C_{AC}C_{AD}}=F_{r_{AB}}^{C_{AB}}\cap
F_{r_{AC}}^{C_{AC}}\cap F_{r_{AD}}^{C_{AD}}$. Clearly, the subfamilies $%
F_{r_{AB}r_{AC}r_{AD}}^{C_{AB}C_{AC}C_{AD}}$ and $F_{r_{AB}^{\prime
}r_{AC}^{\prime }r_{AD}^{\prime }}^{C_{AB}C_{AC}C_{AD}}$ are SLOCC
inequivalent when $r_{AB}r_{AC}r_{AD}\neq r_{AB}^{\prime }r_{AC}^{\prime
}r_{AD}^{\prime }$.

We now further partition the nine families introduced by Verstraete \emph{et al.}
\cite{Verstraete} into SLOCC inequivalent subfamilies via the rank of coefficient
matrix. For convenience, we rewrite the families $G_{abcd}$ and $L_{abc_{2}}$
as:
\begin{eqnarray}
G_{abcd} &=&\alpha (|0\rangle +|15\rangle )+\beta (|3\rangle +|12\rangle
)+\gamma (|5\rangle +|10\rangle )  \notag  \label{eq_Gabcd} \\
&&+\delta (|6\rangle +|9\rangle ), \\
L_{abc_{2}} &=&\alpha ^{\prime }(|0\rangle +|15\rangle )+\beta ^{\prime
}(|3\rangle +|12\rangle) +\gamma ^{\prime }(|5\rangle +|10\rangle )  \notag \\
&&+|6\rangle .  \label{eq_Labc2}
\end{eqnarray}

In Table \ref{table4}, we show the subfamilies $F_{r_{AB}}^{C_{AB}}$, $%
F_{r_{AC}}^{C_{AC}}$, and $F_{r_{AD}}^{C_{AD}}$ of $G_{abcd}$. As
illustrated in Table \ref{table5}, $G_{abcd}$ can be further partitioned
into nine genuinely entangled subfamilies and three biseparable subfamilies
(marked with ``*") via $r_{AB}$, $r_{AC}$, and $r_{AD}$ (subfamilies not
listed in the table are empty). For simplicity, the detailed descriptions of
the subfamilies are not shown as they can be easily obtained by taking the
intersections of the corresponding descriptions in Table \ref{table4}.
Tables \ref{table6} and \ref{table7} illustrate the partitions of the other
eight families introduced by Verstraete \emph{et al.} into inequivalent
subfamilies. In total, we find 28 genuinely entangled subfamilies and all
the degenerate classes can be distinguished up to permutations of the four
qubits (i.e., $A$-$B$-$C$-$D$, $A$-$B$-$CD$, $AB$-$CD$, $|0\rangle_A|W%
\rangle_{BCD}$, and $|0\rangle_A|\mbox{GHZ}\rangle_{BCD}$).

\begin{table*}[tbp]
\caption{The subfamilies $F_{r_{AB}}^{C_{AB}}$, $F_{r_{AC}}^{C_{AC}}$, and $%
F_{r_{AD}}^{C_{AD}}$ of $G_{abcd}$.}
\label{table4}
. \renewcommand\arraystretch{1.2}
\begin{ruledtabular}
\begin{tabular}{ll}
 Subfamily & Description\\ \hline
 $F_{1}^{C_{AB}}$ & $\alpha =\beta =0\ \& \ \gamma =\pm \delta \neq 0\ |\
\alpha=\pm \beta \neq 0\ \&\ \gamma =\delta =0$ \\
 $F_{2}^{C_{AB}}$ & $\alpha =\beta =0\ \&\ \gamma \neq \pm \delta \ |\
\gamma =\delta =0\ \&\ \alpha\neq \pm \beta \ |\
\alpha =\pm \beta \neq 0\ \&\ \gamma =\pm \delta \neq 0$\\
 $F_{3}^{C_{AB}}$ & $\alpha =\pm \beta \neq 0\ \&\
\gamma \neq \pm \delta \ |\ \gamma =\pm \delta \neq 0\ \&\
\alpha \neq \pm \beta$\\
 $F_4^{C_{AB}}$ & $\alpha \neq \pm \beta \ \&\ \gamma \neq \pm \delta$\\
 $F_1^{C_{AC}}$ & $\alpha =\gamma =0 \ \&\ \beta =\pm \delta \neq 0\ |\
\alpha=\pm \gamma \neq 0\ \&\ \beta =\delta =0$\\
 $F_2^{C_{AC}}$ & $\alpha =\gamma =0\ \&\ \beta \neq \pm \delta \ |\
\beta =\delta =0\ \&\ \alpha\neq \pm \gamma \ |\ \alpha =\pm \gamma \neq 0
\ \&\ \beta =\pm \delta \neq 0$\\
 $F_3^{C_{AC}}$ & $\alpha =\pm \gamma \neq 0\ \&\ \beta \neq \pm \delta \ |\
\beta =\pm \delta \neq 0\ \&\ \alpha \neq \pm \gamma$\\
 $F_4^{C_{AC}}$ & $\alpha \neq \pm \gamma \ \&\ \beta \neq \pm \delta$ \\
 $F_1^{C_{AD}}$ & $\alpha =\delta =0\ \&\ \beta =\pm \gamma \neq 0\ |\
\alpha=\pm \delta \neq 0\ \&\ \beta =\gamma =0$\\
 $F_2^{C_{AD}}$ & $\alpha =\delta =0\ \&\ \beta \neq \pm \gamma \ |\
\beta =\gamma =0\ \&\ \alpha\neq \pm \delta \ |\
\alpha =\pm \delta \neq 0\ \&\ \beta =\pm \gamma \neq 0$\\
 $F_3^{C_{AD}}$ & $\alpha =\pm \delta \neq 0\ \&\ \beta \neq \pm \gamma \ |\
\beta =\pm \gamma\neq 0\ \&\ \alpha \neq \pm \delta$\\
 $F_4^{C_{AD}}$ & $\alpha \neq \pm \delta \ \&\ \beta \neq \pm \gamma$
\end{tabular}
\end{ruledtabular}
\end{table*}

\begin{table}[tbp]
\caption{SLOCC classification of $G_{abcd}$ via $r_{AB}$, $r_{AC}$, and $%
r_{AD}$. The subfamilies marked with ``*" are biseparable.}
\label{table5}
. \renewcommand\arraystretch{1.2}
\begin{ruledtabular}
\begin{tabular}{ll}
 $r_{AB}$ $r_{AC}$ $r_{AD}$ & Subfamily description\\ \hline
             222 & $F_{2}^{C_{AB}}\cap F_{2}^{C_{AC}}\cap F_{2}^{C_{AD}} $ \\
             244 & $F_{2}^{C_{AB}}\cap F_{4}^{C_{AC}}\cap F_{4}^{C_{AD}}$ \\
             333 & $F_{3}^{C_{AB}}\cap F_{3}^{C_{AC}}\cap F_{3}^{C_{AD}}$ \\
             344 & $F_{3}^{C_{AB}}\cap F_{4}^{C_{AC}}\cap F_{4}^{C_{AD}}$  \\
             424 & $F_{4}^{C_{AB}}\cap F_{2}^{C_{AC}}\cap F_{4}^{C_{AD}}$ \\
             434 & $F_{4}^{C_{AB}}\cap F_{3}^{C_{AC}}\cap F_{4}^{C_{AD}}$ \\
             442 & $F_{4}^{C_{AB}}\cap F_{4}^{C_{AC}}\cap F_{2}^{C_{AD}}$  \\
             443 & $F_{4}^{C_{AB}}\cap F_{4}^{C_{AC}}\cap F_{3}^{C_{AD}}$\\
             444 & $F_{4}^{C_{AB}}\cap F_{4}^{C_{AC}}\cap F_{4}^{C_{AD}}$  \\
             144$^*$ & $F_{1}^{C_{AB}}$ (i.e.,  $AB$-$CD$) \\
             414$^*$ & $F_{1}^{C_{AC}}$ (i.e., $AC$-$BD$) \\
	     441$^*$ & $F_{1}^{C_{AD}}$ (i.e., $AD$-$BC$) \\
\end{tabular}
\end{ruledtabular}
\end{table}

\begin{table*}[tbp]
\caption{SLOCC classification of $L_{abc_{2}}$ via $r_{AB}$, $r_{AC}$, and $%
r_{AD}$. The subfamilies marked with \textquotedblleft *" are biseparable.}
\label{table6}
. \renewcommand\arraystretch{1.2}
\begin{ruledtabular}
\begin{tabular}{ll}
 $r_{AB}$ $r_{AC}$ $r_{AD}$ & Subfamily description\\ \hline
	     233 & $\alpha' =\beta' =0\ \&\ \gamma' \neq 0$ \\
             244 & $\alpha' =\pm \beta' \neq 0\ \&\ \gamma' =0$ \\
             323 & $\alpha' =\gamma' =0\ \&\ \beta' \neq 0$ \\
             332 & $\alpha' \neq 0\ \&\ \beta' =\gamma' =0$ \\
             333 & $\alpha' =\pm \beta' =\pm \gamma' \neq 0$ \\
             344 & $\gamma' =0\ \&\ \alpha' \beta' \neq 0\ \&\
                   \alpha' \neq \pm \beta' \ |\ \gamma' \neq 0\ \&\
                   \alpha' =\pm \beta' \neq 0\ \&\ \alpha' \neq \pm \gamma'$ \\
             424 & $\beta' =0\ \&\ \alpha' =\pm \gamma' \neq 0$ \\
             434 & $\beta' =0\ \&\ \alpha' \gamma' \neq 0\ \&\
                   \alpha' \neq \pm \gamma' \ |\ \beta' \neq 0 \ \& \
                   \alpha' =\pm \gamma' \neq 0\ \&\ \alpha' \neq \pm \beta'$ \\
             442 & $\alpha' =0\ \&\ \beta' =\pm \gamma' \neq 0$ \\
             443 & $\alpha' =0\ \&\ \beta' \neq \pm \gamma' \ \&\
                   \beta' \gamma' \neq 0 \ |\ \alpha' \neq 0\ \&\
                   \beta' =\pm \gamma' \neq 0\ \&\ \alpha' \neq \pm \beta'$ \\
             444 & $\gamma' \neq 0\ \&\ \alpha' \neq \pm \beta' \ \&\
                   \beta' \neq 0\ \&\ \alpha' \neq \pm \gamma' \ \&\
                   \alpha' \neq 0\ \&\ \beta' \neq \pm \gamma'$ \\
             111$^*$ & $\alpha' =\beta' =\gamma' =0$, (i.e., $A$-$B$-$C$-$D$)
\end{tabular}
\end{ruledtabular}
\end{table*}

\begin{table*}[tbp]
\caption{SLOCC classifications of $L_{ab_{3}}$, $L_{a_{2}b_{2}}$, $L_{a_{4}}$%
, $L_{a_{2}0_{3\oplus {\bar{1}}}}$, $L_{0_{5\oplus {\bar{3}}}}$, $%
L_{0_{7\oplus {\bar{1}}}}$, and $L_{0_{3\oplus {\bar{1}}}0_{3\oplus {\bar{1}}%
}}$ via $r_{AB}$, $r_{AC}$, and $r_{AD}$. The subfamilies marked with
\textquotedblleft *" are biseparable.}
\label{table7}
. \renewcommand\arraystretch{1.2}
\begin{ruledtabular}
\begin{tabular}{llllll}
 Family & $r_{AB}$ $r_{AC}$ $r_{AD}$ & Subfamily description & Family &
$r_{AB}$ $r_{AC}$ $r_{AD}$ & Subfamily description\\ \hline
 $L_{a_{2}b_{2}}$ & 333$$ & $ab=0\ \&\ a\neq b$ &
 $L_{ab_{3}}$
              & 222$$ & $a=b=0$ (i.e., $|W\rangle_{ABCD}$) \\
                                   & 424$$ & $a=\pm b\neq 0$ &
              & 344$$ & $ab=0\ \& \ a\neq b$ \\
                                   & 434 & $ab\neq 0\ \&\ a\neq \pm b$ &
              & 424$$ & $ a=b\neq 0$ \\
                                   & 212$^*$ & $a=b=0$ (i.e., $A$-$C$-$BD$) &
              & 434$$ & $ b=-3a\neq 0$ \\
 $L_{a_{4}}$ & 323$$ & $L_{a_{4}}(a=0)$ &
     & 442$$ & $ a=-b\neq 0$ \\
                              & 434$$ & $L_{a_{4}}(a\neq 0)$ &
              & 443$$ & $ b=3a\neq 0$ \\
 $L_{a_{2}0_{3\oplus {\bar 1}}}$ & 333 & $L_{a_{2}0_{3\oplus{\bar 1}}}(a\neq 0)$ &
              & 444 & $ab\neq 0\ \& \ b\neq \pm a\ \& \ b\neq \pm 3a$ \\
                                 & 222$^*$ & $a=0$ (i.e., $|0\rangle_{A}|W\rangle_{BCD}$) &
 $L_{0_{5\oplus {\bar 3}}}$$$ &333 &  $L_{0_{5\oplus {\bar 3}}}$ \\
 $L_{0_{7\oplus {\bar 1}}}$$$ &333 &   $L_{0_{7\oplus {\bar 1}}}$ &
 $L_{0_{3\oplus {\bar 1}}0_{3\oplus {\bar 1}}}$$$ &222$^*$&  $|0\rangle_{A}
|\mbox{GHZ}\rangle_{BCD}$    \\
\end{tabular}
\end{ruledtabular}
\end{table*}

\section{VI. Discussion of the completeness of the nine families obtained 
by Verstraete \emph{et al.}}

The family $L_{ab_{3}}$ in Ref. \cite{Verstraete} was defined as
\begin{eqnarray}
L_{ab_{3}} &=&a(|0000\rangle +|1111\rangle )+\frac{a+b}{2}(|0101\rangle
+|1010\rangle )  \notag \\
&&+\frac{a-b}{2}(|0110\rangle +|1001\rangle )  \notag \\
&&+\frac{i}{\sqrt{2}}(|0001\rangle +|0010\rangle +|0111\rangle +|1011\rangle
).
\end{eqnarray}%
In later work, Chterental \emph{et al.} \cite{Verstraete} obtained nine SLOCC
inequivalent families of four qubits using invariant theory. Let $%
L_{ab_{3}}^{\prime }$ be defined by
\begin{eqnarray}
L_{ab_{3}}^{\prime } &=&a(|0000\rangle +|1111\rangle )+\frac{a+b}{2}%
(|0101\rangle +|1010\rangle )  \notag \\
&&+\frac{a-b}{2}(|0110\rangle +|1001\rangle )  \notag \\
&&+\frac{i}{\sqrt{2}}(|0001\rangle +|0010\rangle -|0111\rangle -|1011\rangle
),
\end{eqnarray}%
that is, $L_{ab_{3}}^{\prime }$ is obtained by replacing the two
\textquotedblleft +" signs of the last two terms in the formula of $%
L_{ab_{3}}$ by \textquotedblleft -" signs \cite{Chterental}. It is claimed
that there is a perfect correspondence between the nine families obtained by
Verstraete \emph{et al.} (with $L_{ab_{3}}$ replaced by $L_{ab_{3}}^{\prime }$) and
the nine families obtained by Chterental \emph{et al.} \cite{Chterental}. Note that
the formula of $L_{ab_{3}}^{\prime }$ has also been adopted in Ref. \cite{Borsten}%
. Since both Verstraete \emph{et al.} and Chterental \emph{et al.} claimed that the nine
families obtained in their work are inequivalent to each other, a detailed
study of the relation between $L_{ab_{3}}$ and $L_{ab_{3}}^{\prime }$ can
provide insights into the completeness of their classifications.

\subsection{A. $L_{ab_{3}}(a=0)$ is SLOCC equivalent to $L_{ab_{3}}^{\prime
}(a=0)$}

It is readily verified that the following equation holds between $%
L_{ab_{3}}^{\prime }(a=0)$ and $L_{ab_{3}}(a=0)$:
\begin{equation}
L_{ab_{3}}^{\prime }(a=0)=I\otimes I\otimes i\sigma _{z}\otimes i\sigma
_{z}L_{ab_{3}}(a=0),  \label{eq-1}
\end{equation}%
where $I$ is the identity and $\sigma _{z}=\mbox{diag}\{1,-1\}$.

It follows from Eq. (\ref{eq-1}) that $L_{ab_{3}}(a=0)$ and $%
L_{ab_{3}}^{\prime }(a=0)$ are SLOCC equivalent. In particular, setting $b=0$
yields that the states $\frac{i}{\sqrt{2}}(|0001\rangle +|0010\rangle
-|0111\rangle -|1011\rangle )$ and $\frac{i}{\sqrt{2}}(|0001\rangle
+|0010\rangle +|0111\rangle +|1011\rangle )$ are equivalent under SLOCC.

\begin{table}[tbp]
\caption{SLOCC classification of $L_{ab_{3}}^{\prime }$ via $r_{AB}$, $%
r_{AC} $, and $r_{AD}$.}
\label{table8}
. \renewcommand\arraystretch{1.2}
\begin{ruledtabular}
\begin{tabular}{ll}
  $r_{AB}$ $r_{AC}$ $r_{AD}$ & Subfamily description \\ \hline
               222 & $a=b=0$ (i.e., $|W\rangle_{ABCD}$) \\
               344 & $ab=0\ \& \ a\neq b$\\
               424& $\emptyset$ \\
               434 & $ a=b\neq 0 \ |\  b=-3a\neq 0$ \\
               442 & $\emptyset$ \\
               443 & $ a=-b\neq 0 \ |\  b=3a\neq 0$ \\
               444 & $ ab\neq 0\ \& \ b\neq \pm a\ \& \ b\neq \pm 3a$ \\
\end{tabular}
\end{ruledtabular}
\end{table}

\subsection{B. $L_{ab_{3}}^{\prime }(a\neq 0)$ [respectively, $L_{ab_{3}}(a\neq 0)$] is
SLOCC inequivalent to $L_{ab_{3}}$ (respectively, $L_{ab_{3}}^{\prime }$)}

We first show that the family $L_{ab_{3}}^{\prime }(a\neq 0)$ is SLOCC
inequivalent to the family $L_{ab_{3}}$. In Table \ref{table8} we show the
partition of $L_{ab_{3}}^{\prime }$ into SLOCC inequivalent subfamilies via $%
r_{AB}$, $r_{AC}$, and $r_{AD}$. Consulting Tables \ref{table7} and \ref%
{table8}, and using the fact that the subfamilies with different ranks of
coefficient matrices are SLOCC inequivalent to each other, it suffices to
consider the following six cases.

{\center \quad \sl{Case 1.} $L_{ab_{3}}^{\prime }(a=b\neq 0)$ is SLOCC inequivalent to $L_{ab_{3}}(b=-3a\neq 0)$.} \newline

In this case, we can resort to $D_{xy}$, a degree 6 polynomial invariant of
four qubits \cite{Luque} (see the Appendix for the expression of $D_{xy}$).
Indeed, it can be verified that if $|\psi \rangle $ and $|\psi ^{\prime
}\rangle $ are any two SLOCC equivalent states, that is, they satisfy Eq. (\ref%
{slocc-eq}), then the following equation holds:
\begin{equation}
D_{xy}(|\psi ^{\prime }\rangle )=D_{xy}(|\psi \rangle )\big[\Pi
_{i=1}^{4}\det \mathcal{A}_{i}\bigr]^{3}.  \label{det-1}
\end{equation}%
It follows from Eq. (\ref{det-1}) that for any two SLOCC equivalent states $%
|\psi \rangle $ and $|\psi ^{\prime }\rangle $, either $D_{xy}(|\psi
^{\prime }\rangle )$ and $D_{xy}(|\psi \rangle )$ both vanish or neither
vanishes.

A direct calculation shows that
\begin{equation}
D_{xy}=-\frac{1}{32}\left( a-b\right) ^{3}\left( a+b\right) ^{3}
\label{inv-1}
\end{equation}%
for both $L_{ab_{3}}$ and $L_{ab_{3}}^{\prime }$. The desired result then
follows by noting that $D_{xy}=16a^{6}\neq 0$ for $L_{ab_{3}}(b=-3a\neq 0)$
whereas $D_{xy}=0$ for $L_{ab_{3}}^{\prime }(a=b\neq 0)$.

{\center \quad \sl{Case 2.} $L_{ab_{3}}^{\prime }(a=-b\neq 0)$ is SLOCC inequivalent to
$L_{ab_{3}}(b=3a\neq 0)$.} \newline

This case can be dealt with similarly as case 1 by noting that $%
D_{xy}=16a^{6}\neq 0$ for $L_{ab_{3}}(b=3a\neq 0)$ whereas $D_{xy}=0$ for $%
L_{ab_{3}}^{\prime }(a=-b\neq 0)$.

{\center \quad \sl{Case 3.} $L_{ab_{3}}^{\prime }(b=-3a\neq 0)$ is SLOCC inequivalent
to $L_{ab_{3}}(b=-3a\neq 0)$.} \newline

In this case, the semi-invariants defined in Ref. \cite{LDF07b} turn out to be
useful. More specifically, for any four-qubit state $|\psi
\rangle=\sum_{i=0}^{15}c_{i}|i\rangle$, the semi-invariants $F_1$ and $F_2$
are defined in Ref. \cite{LDF07b} as
\begin{eqnarray}
F_{1}(\psi)&=&(c_{0}c_{7}-c_{2}c_{5}+c_{1}c_{6}-c_{3}c_{4})^{2}  \notag \\
& & -4(c_{2}c_{4}-c_{0}c_{6})(c_{3}c_{5}-c_{1}c_{7}),
\label{eq_semiinvariant1} \\
F_{2}(\psi)&=&(c_{8}c_{15}-c_{11}c_{12}+c_{9}c_{14}-c_{10}c_{13})^{2}  \notag
\\
& & -4(c_{11}c_{13}-c_{9}c_{15})(c_{10}c_{12}-c_{8}c_{14}).
\label{eq_semiinvariant2}
\end{eqnarray}

Let $|\phi \rangle $\ be any four-qubit state SLOCC equivalent to $%
L_{ab_{3}} $ [i.e., they satisfy Eq. (\ref{slocc-eq})]. Let
\begin{equation}
\mathcal{A}_{1}=\left(
\begin{array}{cc}
\alpha _{1} & \alpha _{2} \\
\alpha _{3} & \alpha _{4}%
\end{array}%
\right) .  \label{def_A}
\end{equation}%
A tedious but straightforward calculation yields
\begin{eqnarray}
F_{1}(\phi ) &=&\frac{1}{2}(a^{2}-b^{2})\alpha _{1}^{4}\biggl[%
\prod_{i=2}^{4}\det \mathcal{A}_{i}\biggr]^{2},  \label{semi-inv-1} \\
F_{2}(\phi ) &=&\frac{1}{2}(a^{2}-b^{2})\alpha _{3}^{4}\biggl[%
\prod_{i=2}^{4}\det \mathcal{A}_{i}\biggr]^{2}.  \label{semi-inv-2}
\end{eqnarray}%
In view of Eqs. (\ref{semi-inv-1}) and (\ref{semi-inv-2}) and the fact that $%
\mathcal{A}_{1}$ is invertible, it follows at once that if $|\phi \rangle $
is SLOCC equivalent to $L_{ab_{3}}(a\neq \pm b)$, then the following
equation holds:
\begin{equation}
\left\vert F_{1}(\phi )\right\vert +\left\vert F_{2}(\phi )\right\vert \neq
0.  \label{relation-1}
\end{equation}

Let $|\varphi \rangle $ be any state SLOCC equivalent to $L_{ab_{3}}^{\prime
}$ [i.e., they satisfy Eq. (\ref{slocc-eq})]. Again, a tedious but
straightforward calculation yields
\begin{eqnarray}
F_{1}(\varphi ) &=&\frac{-1}{2\sqrt{2}}i\alpha _{1}^{3}\bigl(-i\sqrt{2}%
(3a^{2}+b^{2})\alpha _{1}+8a(a^{2}-b^{2})\alpha _{2}\bigr)  \notag \\
&&\times \biggl[\prod_{i=2}^{4}\det \mathcal{A}_{i}\biggr]^2,
\label{semi-inv-3} \\
F_{2}(\varphi ) &=&\frac{-1}{2\sqrt{2}}i\alpha _{3}^{3}\bigl(-i\sqrt{2}%
(3a^{2}+b^{2})\alpha _{3}+8a(a^{2}-b^{2})\alpha _{4}\bigr)  \notag \\
&&\times \biggl[\prod_{i=2}^{4}\det \mathcal{A}_{i}\biggr]^2.
\label{semi-inv-4}
\end{eqnarray}

When $a(a^{2}-b^{2})\neq 0$,\ consider the operator
\begin{equation}  \label{eq_A}
\mathcal{A}_{1}^{\ast }=\left(
\begin{array}{cc}
\alpha _{1} & \frac{i\sqrt{2}(3a^{2}+b^{2})}{8a(a^{2}-b^{2})}\alpha _{1} \\
0 & \alpha _{4}%
\end{array}%
\right),
\end{equation}
where $\alpha _{1}\alpha _{4}\neq 0$. Clearly, $\mathcal{A}_{1}^{\ast}$ is
invertible. In view of Eqs. (\ref{semi-inv-3})-(\ref{eq_A}), 
it follows that there exists a state $|\varphi ^{\ast }\rangle $
equivalent to $L_{ab_{3}}^{\prime }(a(a^{2}-b^{2})\neq 0)$ under local
invertible operators $\mathcal{A}_{1}^{\ast }$, $\mathcal{A}_{2}$, $\mathcal{%
A}_{3}$, and $\mathcal{A}_{4}$, such that
\begin{equation}
\left\vert F_{1}(\varphi ^{\ast })\right\vert +\left\vert F_{2}(\varphi
^{\ast })\right\vert =0.  \label{relation-2}
\end{equation}%
From Eqs. (\ref{relation-1}) and (\ref{relation-2}), $|\varphi ^{\ast
}\rangle $\ is SLOCC inequivalent to the state $L_{ab_{3}}(a\neq \pm b)$.\
Therefore, $L_{ab_{3}}^{\prime }(a(a^{2}-b^{2})\neq 0)$ is SLOCC
inequivalent to $L_{ab_{3}}(a\neq \pm b)$. In particular, $%
L_{ab_{3}}^{\prime }(b=-3a\neq 0)$ is SLOCC inequivalent to $%
L_{ab_{3}}(b=-3a\neq 0)$.

{\center \quad \sl{Case 4.} $L_{ab_{3}}^{\prime }(b=3a\neq 0)$ is SLOCC inequivalent to
$L_{ab_{3}}(b=3a\neq 0)$.} \newline

This case can be treated analogously to case 3.

{\center \quad \sl{Case 5.} $L_{ab_{3}}^{\prime }(a\neq 0\ \&\ b=0)$ is SLOCC
inequivalent to $L_{ab_{3}}(ab=0\ \&\ a\neq b)$.} \newline

In Ref. \cite{LDFQIC09}, we proved that $L_{ab_{3}}(a=0\ \&\ b\neq 0)$ and $%
L_{ab_{3}}(a\neq 0\ \&\ b=0)$ are SLOCC inequivalent. A proof analogous to
that of Ref. \cite{LDFQIC09} shows that $L_{ab_{3}}^{\prime }(a=0\ \&\ b\neq 0)$
and $L_{ab_{3}}^{\prime }(a\neq 0\ \&\ b=0)$ are SLOCC inequivalent. Using
the fact that $L_{ab_{3}}(a=0\ \&\ b\neq 0)$ is SLOCC equivalent to $%
L_{ab_{3}}^{\prime }(a=0\ \&\ b\neq 0)$ [see Eq. (\ref{eq-1})] yields that $%
L_{ab_{3}}^{\prime }(a\neq 0\ \&\ b=0)$ is SLOCC inequivalent to $%
L_{ab_{3}}(a=0\ \&\ b\neq 0)$. Furthermore, an argument analogous to case 3
shows that $L_{ab_{3}}^{\prime }(a\neq 0\ \&\ b=0)$ is inequivalent to $%
L_{ab_{3}}(a\neq 0\ \&\ b=0)$.

Indeed, we can further conclude that $L_{ab_{3}}(a=0)$ and $L_{ab_{3}}(a\neq
0)$ are SLOCC inequivalent and $L_{ab_{3}}^{\prime }(a=0)$ and $%
L_{ab_{3}}^{\prime }(a\neq 0)$ are SLOCC inequivalent.

{\center \quad \sl{Case 6.} $L_{ab_{3}}^{\prime }(ab\neq 0\ \&\ a\neq \pm b\ \&\ b\neq
\pm 3a)$ is SLOCC inequivalent to $L_{ab_{3}}(ab\neq 0\ \&\ a\neq \pm b\ \&\
b\neq \pm 3a)$.} \newline

This case can be treated analogously to case 3.

As a consequence, $L_{ab_{3}}^{\prime }(a\neq 0)$ is SLOCC inequivalent to $%
L_{ab_{3}}$. An analogous argument shows that $L_{ab_{3}}(a\neq 0)$ is SLOCC
inequivalent to $L_{ab_{3}}^{\prime }$.

\subsection{C. The relation between $L_{ab_{3}}^{\prime }\ $and $L_{ab_{3}}$
under permutations}

Let $|\gamma \rangle $ be the state of the subfamily $L_{ab_{3}}^{\prime
}(a\neq 0\ \&\ b=0)$, $|\eta \rangle $ be the state of the subfamily $%
L_{ab_{3}}^{\prime }(b=3a\neq 0)$, $|\vartheta \rangle $ be the state of the
subfamily $L_{ab_{3}}^{\prime }(b=-3a\neq 0)$, and $|\nu \rangle $ be the
state of the subfamily $L_{ab_{3}}^{\prime }(ab\neq 0\ \&\ a \neq \pm b\ \&\ 
b\neq \pm 3a)$. We argue that the above four subfamilies are SLOCC inequivalent
to $L_{ab_{3}}$ under any permutation of qubits. This can be seen as follows.
Let $(i,j)$ be the transposition of qubits $i$ and $j$. A tedious calculation 
shows that 
the permutations giving rise to different $|\gamma \rangle $ are $\kappa
_{1}=I$, $\kappa _{2}=(1,3)$, $\kappa _{3}=(1,4)$, $\kappa _{4}=(1,2)(1,3)$,
$\kappa _{5}=(1,2)(1,4)$, and $\kappa _{6}=(1,4)(1,2)(1,3)$. Similarly, 
the permutations giving rise to different $|\eta\rangle $, 
$|\vartheta \rangle$, and $|\nu \rangle $ are
$\pi _{1}=I$, $\pi _{2}=(1,2)$, $\pi _{3}=(1,3)$%
, $\pi _{4}=(1,4)$, $\pi _{5}=(1,3)(1,2)$, $\pi _{6}=(1,4)(1,2)$, $\pi
_{7}=(1,2)(1,3)$, $\pi _{8}=(1,2)(1,4)$, $\pi _{9}=(1,2)(1,3)(1,2)$, $\pi
_{10}=(1,2)(1,4)(1,2)$, $\pi _{11}=(1,4)(1,2)(1,3)$, and $\pi
_{12}=(1,4)(1,2)(1,3)(1,2)$. 
The result that $\kappa _{i}|\gamma \rangle (i=1,\cdots, 6)$, 
$\pi_{j}|\eta \rangle $, $\pi_{j}|\vartheta \rangle $, and 
$\pi_{j}|\nu \rangle (j=1,\cdots, 12)$ are all SLOCC inequivalent to $L_{ab_{3}}$ 
then follows by calculating the ranks $r_{AB}$, $r_{AC}$, and $r_{AD}$
of $\kappa _{i}|\gamma \rangle $, $\pi _{j}|\eta \rangle $, $\pi
_{j}|\vartheta \rangle $ and $\pi _{j}|\nu \rangle $, and using 
an argument analogous to that of case 3 in the previous section.

\textsl{Remark 2. }By using Tables \ref{table7} and \ref{table8}, one can 
verify that $(1,4)L_{ab_{3}}^{\prime }(a=b\neq 0)$ is SLOCC equivalent to 
$L_{ab_{3}}(a=0\ \&\ b\neq 0)$ under the invertible local operator 
$\sigma_{x}\otimes \sigma _{z}\otimes iI\otimes \sigma _{y}$, and 
$(1,3)L_{ab_{3}}^{\prime }(a=-b\neq 0)$ is SLOCC equivalent to 
$L_{ab_{3}}(a=0\ \&\ b\neq 0)$ under the invertible local operator 
$\sigma_{x}\otimes \sigma _{z}\otimes \sigma _{y}\otimes iI$.

\subsection{D. $L_{ab_{3}}^{\prime }(a\neq 0)$ is SLOCC inequivalent to the
other eight families by Verstraete \emph{et al.}}

Here we show that $L_{ab_{3}}^{\prime }(a\neq 0)$ is not only SLOCC
inequivalent to $L_{ab_{3}}$ but also SLOCC inequivalent to the other eight
families by Verstraete \emph{et al.} For simplicity, we only show that $%
L_{ab_{3}}^{\prime }(a=-b\neq 0)$ is SLOCC inequivalent to the other eight
families obtained by Verstraete \emph{et al.} From Table \ref{table8}, $%
r_{AB}r_{AC}r_{AD}=443$ for $L_{ab_{3}}^{\prime }(a=-b\neq 0)$. Consulting
Tables \ref{table5}, \ref{table6}, and \ref{table7}, and using the fact that
the subfamilies with different ranks of coefficient matrices are SLOCC
inequivalent to each other, it suffices to show that $L_{ab_{3}}^{\prime
}(a=-b\neq 0)$ is SLOCC inequivalent to the subfamilies with $%
r_{AB}r_{AC}r_{AD}=443$ of $G_{abcd}$ and $L_{abc_{2}}$.

To show that $L_{ab_{3}}^{\prime }(a=-b\neq 0)$ is SLOCC inequivalent to the
subfamily with $r_{AB}r_{AC}r_{AD}=443$ of $G_{abcd}$, we use the degree 6
polynomial invariant $D_{xy}$ given in Eq. (\ref{det-1}). It is readily seen
from Eq. (\ref{inv-1}) that $D_{xy}=0$ for $L_{ab_{3}}^{\prime }(a=-b\neq 0)$%
. A simple calculation shows that
\begin{equation}
D_{xy}=(\alpha \beta -\gamma \delta )(\alpha \beta +\gamma \delta )(\alpha
^{2}+\beta ^{2}-\gamma ^{2}-\delta ^{2})  \label{Dxy-2}
\end{equation}%
for $G_{abcd}$ [as defined in Eq. (\ref{eq_Gabcd})]. It is readily seen from
Eq. (\ref{Dxy-2}) that $D_{xy}\neq 0$ for the subfamily with $%
r_{AB}r_{AC}r_{AD}=443$ of $G_{abcd}$ and then the desired result follows.

Next we show that $L_{ab_{3}}^{\prime }(a=-b\neq 0)$ is SLOCC inequivalent
to the subfamily with $r_{AB}r_{AC}r_{AD}=443$ of $L_{abc_{2}}$ [as defined
in Eq. (\ref{eq_Labc2})]. A calculation shows that
\begin{equation}
D_{xy}=(\alpha^{\prime }\beta^{\prime })^{2}(\alpha^{\prime 2}-\gamma
^{\prime 2}+\beta ^{\prime 2})  \label{Dxy-4}
\end{equation}%
for $L_{abc_{2}}$. From Table \ref{table6}, 
we distinguish the following two cases.

{\center \quad \sl{Case 1.} $\alpha^{\prime }\neq 0\ \&\ \beta ^{\prime } =\pm \gamma ^{\prime}\neq 0\ \&\ \alpha^{\prime }\neq \pm \beta ^{\prime }$.}

In this case $D_{xy}\neq 0$ and then the desired result follows.

{\center \quad \sl{Case 2.} $\alpha^{\prime }=0\ \&\ \beta ^{\prime }\neq \pm \gamma^{\prime } \ \&\ \beta^{\prime }\gamma ^{\prime }\neq 0$.}

In this case $D_{xy}=0$. We can resort to the semi-invariants given in Eqs. (%
\ref{eq_semiinvariant1}) and (\ref{eq_semiinvariant2}). Let $|\varphi\rangle
$ be any state SLOCC equivalent to $L_{ab_{3}}^{\prime }(a=-b\neq 0)$ with 
$\mathcal{A}_1$ given by Eq. (\ref{def_A}). 
A tedious but straightforward calculation yields
\begin{eqnarray}
F_{1}(|\varphi\rangle ) &=&-2a^{2}\alpha _{1}^{4}\biggl[\prod _{i=2}^{4}\det
\mathcal{A}_{i}\biggr]^{2},  \label{F-1} \\
F_{2}(|\varphi\rangle ) &=&-2a^{2}\alpha _{3}^{4}\biggl[\prod _{i=2}^{4}\det
\mathcal{A}_{i}\biggr]^{2}.  \label{F-2}
\end{eqnarray}%
In view of Eqs. (\ref{F-1}) and (\ref{F-2}) and the fact that $\mathcal{A}%
_{1}$ is invertible, it follows at once that if $|\varphi\rangle$ is SLOCC
equivalent to $L_{ab_{3}}^{\prime }(a=-b\neq 0)$, then the following
equation holds:
\begin{equation}
\left\vert F_{1}(\varphi)\right\vert +\left\vert F_{2}(\varphi)\right\vert
\neq 0.
\end{equation}
The desired result then follows by noting that $F_{1}=F_{2}=0$ for $%
L_{abc_{2}}$ with $\alpha^{\prime }=0\ \&\ \beta^{\prime }\neq \pm
\gamma^{\prime }\ \&\ \beta^{\prime }\gamma^{\prime }\neq 0$.

As a consequence, $L_{ab_{3}}^{\prime }(a=-b\neq 0)$ is SLOCC inequivalent
to the nine families obtained by Verstraete \emph{et al.} \cite{Verstraete}.

The discussion suggests that the partition in Ref. \cite{Verstraete} is incomplete.
For completeness, one may add the family $L_{ab_{3}}^{\prime }$ to the
family $L_{ab_{3}}$ in Ref. \cite{Verstraete}. An analogous argument shows that the
partition in Ref. \cite{Chterental} is incomplete as well, and for completeness,
one may add the family $L_{ab_{3}}$ to the family 6 in Ref. \cite{Chterental}.

\section{VII. Conclusion}

We have recast the necessary and sufficient condition for two $n$-qubit 
states to be equivalent under SLOCC into an equivalent form in terms of the 
coefficient matrices associated with the states. As a direct consequence of 
the new necessary and sufficient condition, we have showed that the rank of the
coefficient matrix as well as the rank of the $\ell$-qubit reduced density 
matrix is invariant under SLOCC. 
We have also presented a recursive formula for the calculation of the rank of 
coefficient matrix of an $n$-qubit biseparable state. The recursive formula 
further gives rise to a biseparability criterion in terms of the rank of 
coefficient matrix to determine if an arbitrary $n$-qubit pure state is 
biseparable. The invariance of the rank of coefficient matrix together with
the biseparability criterion reveals that all the degenerate families of 
general $n$ qubits are inequivalent under SLOCC. 

We have then classified four-qubit states under SLOCC via the ranks of 
coefficient matrices and the nine families introduced by Verstraete \emph{et al.}
were further partitioned into inequivalent subfamilies. 
In particular, we have found 28 genuinely entangled 
families and all the degenerate classes can be distinguished up to permutations
of the four qubits.
We have performed a detailed study of the relation between
the family $L_{ab_{3}}$ and the family $L'_{ab_{3}}$ with corrections 
to the signs of the last two terms in the formula of $L_{ab_{3}}$
via the ranks of coefficient matrices. 
By using a degree 6 polynomial invariant and two semi-invariants of four
qubits, we have found that $L'_{ab_{3}}(a=0)$ is SLOCC equivalent to  
$L'_{ab_{3}}(a=0)$ whereas $L'_{ab_{3}}(a\not =0)$ is SLOCC  
inequivalent to $L_{ab_{3}}(a\not =0)$. 
We have also demonstrated that  $L_{ab_{3}}^{\prime }(a\neq 0\ \&\ b=0)$, 
$L_{ab_{3}}^{\prime }(b=\pm 3a\neq 0)$, and  $L_{ab_{3}}^{\prime }(ab\neq 0\
\&\ a\neq \pm b\ \&\ b\neq \pm 3a)$ are SLOCC inequivalent to $L_{ab_{3}}$
under any permutation of qubits, whereas $L_{ab_{3}}^{\prime }(a=\pm b\neq 0)$
are SLOCC equivalent to $L_{ab_{3}}(a=0\ \&\ b\neq 0)$
under some permutations.
This suggests that the partition of four-qubit states into the nine
families by Verstraete \emph{et al.} is incomplete, and for completeness,
one may simply add the family $L_{ab_{3}}^{\prime }$ to the family $L_{ab_{3}}$.

\section{Acknowledgement}
This work was supported by NSFC Grant No. 10875061 and Tsinghua National
Laboratory for Information Science and Technology.

\section{Appendix}

Following \cite{Luque}, $D_{xy}$ can be constructed as
\begin{equation}
D_{xy}=\left\vert
\begin{array}{ccc}
d_{11} & d_{12} & d_{13} \\
d_{21} & d_{22} & d_{23} \\
d_{31} & d_{32} & d_{33}%
\end{array}%
\right\vert ,
\end{equation}%
where the entries of $D_{xy}$ are given by:
\begin{eqnarray}
d_{11} &=&a_{0}a_{3}-a_{1}a_{2},  \notag \\
d_{12} &=&a_{0}a_{7}-a_{1}a_{6}-a_{2}a_{5}+a_{3}a_{4},  \notag \\
d_{13} &=&a_{4}a_{7}-a_{5}a_{6},  \notag \\
d_{21} &=&a_{0}a_{11}-a_{1}a_{10}-a_{2}a_{9}+a_{3}a_{8},  \notag \\
d_{22} &=&a_{0}a_{15}-a_{1}a_{14}-a_{2}a_{13}+a_{3}a_{12}  \notag \\
&&+a_{4}a_{11}-a_{5}a_{10}-a_{6}a_{9}+a_{7}a_{8}, \\
d_{23} &=&a_{4}a_{15}-a_{5}a_{14}-a_{6}a_{13}+a_{7}a_{12},  \notag \\
d_{31} &=&a_{8}a_{11}-a_{9}a_{10},  \notag \\
d_{32} &=&a_{8}a_{15}-a_{9}a_{14}-a_{10}a_{13}+a_{11}a_{12},  \notag \\
d_{33} &=&a_{12}a_{15}-a_{13}a_{14}.  \notag
\end{eqnarray}

\end{document}